\documentstyle[12pt]{article}
\catcode`\@=11
\@addtoreset{equation}{section}

\def\anp#1#2#3{
        Ann. Phys. (N.Y.) {\bf #1 }, #2 (19#3)}

\def\ibid#1#2#3{
	{\it ibid.} {\bf #1}, #2 (19#3)}
\def\jph#1#2#3{
	Jour. of Phys. {\bf #1}, #2 (19#3)}
\def\phl#1#2#3{
        Phys. Lett. {\bf #1}, #2 (19#3)}
\def\prl#1#2#3{
        Phys. Rev. Lett. {\bf #1}, #2 (19#3)}
\def\rmp#1#2#3{
        Rev. Mod. Phys. {\bf#1}, #2 (19#3)}

\def\prd#1#2#3{
        Phys. Rev. D {\bf #1}, #2 (19#3)}
\def\nup#1#2#3{
        Nucl. Phys. {\bf #1}, #2 (19#3)}
\def\zpc#1#2#3{
        Z. Phys. C {\bf #1}, #2 (19#3)}

\def\beq{\begin{equation}}
\def\eeq{\end{equation}}
\def\zn{$Z(N)\;$}
\def\sun{$SU(N)\;$}
\def\tl{\tilde{\lambda}}
\begin{document}
\begin{titlepage}
\begin{flushright}
BNL--GP--1/93\\
January, 1993\\
\end{flushright}
\vfill
\begin{center}
{\Large \bf Partition function for the eigenvalues of the Wilson line}\\
%
%
\vfill
{\large \bf Andreas Gocksch\\
\& \\
Robert D. Pisarski}\\
$\;$ \\
Department of Physics\\
Brookhaven National Laboratory\\
Upton, New York  11973 \\
\vfill
{\large \bf Abstract}
\end{center}
\begin{quotation}

In a gauge theory at nonzero temperature the
eigenvalues of the Wilson line form a set of
gauge invariant observables.  By constructing
the corresponding
partition function for the phases of these eigenvalues,
we prove that the trivial vacuum, where the phases vanish, is
a minimum of the free energy.

\end{quotation}
\vfill
\end{titlepage}

In computing the properties of gauge theories at a nonzero temperature $T$,
a standard method of calculation is to use the imaginary time formalism,
where the gauge
fields $A_\mu$ are strictly periodic
in a euclidean time $\tau$, with period
$\beta = 1/T$ [\ref{r1}].
This allows for a type of Aharonov-Bohm effect at
nonzero temperature,
characterized by a nonzero expectation value for
the Wilson line that wraps around
in the direction of the imaginary time,
$exp(i g \int A_0 \, d \tau)$.
At one loop order [\ref{r2}-\ref{r4}]
these inequivalent vacua are parametrized simply by
a constant value for the gauge potential $A_0$;
calculation shows that the trivial vacuum, with $A_0 =0$, minimizes
the free energy.  At two loop order [\ref{r5}]
it is necessary to
chararcterize the vacua not just by constant $A_0$, but
by the Wilson line itself [\ref{r6},\ref{r7}];
after doing so, the trivial vacuum remains a minimum.

As conjectured in ref. [\ref{r7}],
in this paper we show that the trivial vacuum is
{\it always} a minimum
of the free energy.
Our starting point is a gauge invariant quantity in euclidean
spacetime, the effective potential for the phases of the eigenvalues of
the Wilson line.  This euclidean path integral is then transformed
into a partition function [\ref{r8}],
where the constraint on the phases of the
eigenvalues becomes an imaginary
chemical potential for global color charge.  Because this
chemical potential is imaginary, it follows directly
that the trival vacuum minimizes the free energy,
up to standard degeneracies.

Typically the partition function of a gauge theory
is a sum not over all states,
but only over those which are
gauge invariant [\ref{r1},\ref{r6},\ref{r9}].  This is true, for example,
in computing the free energy at nonzero temperature.
In our case,
while the {\it sum} over states is gauge invariant, the individual
states which contribute are {\it not}.  In this vein, we note that
while partition functions with a chemical potential for global
charge have been studied previously [\ref{r10}], hitherto it was
assumed that only gauge invariant states contribute.
The extension of what is an allowed partition function is, we feel,
the most striking feature of our results.

For definiteness we consider an \sun gauge theory without matter fields,
following previous conventions [\ref{r7}].  Under a
local gauge transformation
$\Omega$, the gauge potential transforms as
$A_\mu^\Omega = \Omega^\dagger D_\mu \Omega/(-i g)$, with $g$ the
coupling constant, and $D_\mu = \partial_\mu - i g A_\mu$ the covariant
derivative.  The Wilson line in the direction of euclidean time is
\beq
L(x) \; = \; {\cal P} \;
exp\left( i g \int^\beta_0 \; A_0(x,\tau) \; d\tau \right)
\;\; \equiv \;\;  U(x) \, \Lambda(x) \, U^\dagger(x) \; ;
\label{e1}
\eeq
${\cal P}$ denotes path ordering, and $x$ is the coordinate for three
spatial dimensions.  The matrix $L(x)$ is unitary, and
so we can write it as the unitary transformation of a diagonal
matrix, $\Lambda(x)$: $\Lambda_{i j}(x) = \delta_{i j} \Lambda_i(x)$,
$U U^\dagger = 1$.
Under a gauge transformation the Wilson line transforms as
\beq
L(x) \rightarrow \Omega(x,\beta) \; {\cal P} \;
 exp \left(i g \int^\beta_0 \;
A^\Omega_0(x,\tau) \; d \tau \right)  \; \Omega^\dagger(x,0) \; .
\eeq
Without loss of generality we require
that both the gauge fields and the gauge transformations are
strictly periodic in the euclidean time.
If $\Omega(x,\beta) = \Omega(x,0)$, then for the Wilson line a
gauge transformation is just a
similarity transformation; the matrices $U(x)$ are gauge
variant, $U(x) \rightarrow \Omega(x,0) U(x)$, while the diagonal
matrix $\Lambda(x)$ is gauge invariant.

Notice that what we call the Wilson
line is a matrix in color space.  This is is distinct from
what is usually termed
Wilson (or Polyakov) loop at nonzero temperature, which is
the trace of this matrix [\ref{r1}].  Since
$tr \, L = tr \,\Lambda$, the
trace of the Wilson line is automatically gauge invariant.

The phases of the eigenvalues are given by
$\Lambda(x) = e^{i \lambda(x)}$; the $\lambda$ are elements of the
Cartan sub-algebra, which is the set of mutually commuting generators
in the group.  We introduce an effective potential for the phases as
\beq
Z(\tl) \; = \;
\int_{A_\mu(x,\beta) = + A_\mu(x,0)} \;
{\cal D}A_\mu(x,\tau) \;\;e^{-S} \;\;
\delta\left(\tl - \int \frac{d^3 x}{V} \; \lambda(x) \right)
\; .
\label{e2}
\eeq
${\cal D}A_\mu$ is the functional measure for the gauge fields,
including gauge fixing and ghosts, $S$ is the gauge field action,
and $V$ is the volume of space.
We have chosen to use a ``constraint''
effective potential [\ref{r11}]; a more standard form, using
an external source and then Legendre transformation,
could also be used.
{}From $Z(\tl)$ the free energy, as a function of the $\tl$, is
$F(\tl) = - log(Z(\tl))$.

When the phases vanish, $F(0)$
reduces to the usual free energy at a
temperature $T$, in the absence of any background
field.  (This is seen most easily from (\ref{e6}) below.)
It is for this reason that we constrain the phases of the eigenvalues
instead of the eigenvalues themselves.
Since the $\tl$'s are phases,
the free energy is periodic: for $\tl_{i j} = \tl_i \delta_{i j}$,
$F(\tl)$ is unchanged
when any single element is shifted by a multiple of $2\pi$,
$F(\tl_i + 2 \pi) = F(\tl_i)$.

In a pure gauge theory there is a further degeneracy in $F(\tl)$.
The gauge potentials remain periodic under certain
aperiodic gauge transformations, as long as the aperiodicity is a
constant element
in the center of the group [\ref{r1}].
For an \sun gauge group the center is \zn,
and the aperiodic gauge transformations satisfy
$\Omega(x,\beta) = e^{i \theta_\ell} \Omega(x,0)$,
where $\theta_\ell = 2 \pi \ell/N$,
and $\ell$ is an integer $= 0,1,...(N-1)$.
Under such gauge transformations every phase transforms by the
same constant, $\lambda_i(x) \rightarrow
\lambda_i(x) + \theta_\ell$ for all $i=1...N$.
This implies that the effective potential
automatically posseses a \zn symmetry,
$F(\tl + \theta_\ell) = F(\tl)$ (here $\theta_\ell$ denotes a
diagonal matrix where each element $= \theta_\ell$; this notation
is used later, and should be clear from the context).

Before turning the effective potential into a partition function, it is
worth reviewing the results of perturbative
calculations [\ref{r2}-\ref{r7}].  To compute one expands
about a constant, background field
$A_0 = A_0^{cl} + A^{qu}_0$, where $A_0^{cl} = \tl/g$.
A convenient gauge is covariant background field gauge, where
the gauge fixing term in the lagrangian is
\beq
{\cal L}_{covariant} \; = \;
\frac{1}{\xi} \;
tr \left( \left( D_\mu^{cl} A^{qu}_\mu \right)^2 \right) \; ,
\eeq
$D^{cl}_\mu = \partial_\mu - i g [A^{cl}_\mu,]$.
At one loop order the free energy is independent of the gauge fixing
parameter $\xi$, $F_{one \; loop}(\tl)
= 2 \, tr \,log(- (D^{cl}_\mu)^2)$; this
is the free energy for two spatially transverse gluons in the
background field $A_0^{cl}$.  The minimum is at $\tl = 0$,
$F_{one \; loop}(0) \leq F_{one \; loop}(\tl)$, plus \zn
transforms thereof.

At two loop order, as a function of $A_0^{cl}$ the free energy is
gauge dependent.  Belyaev [\ref{r6},\ref{r7}] noted that
this happens
because at one loop order fluctuations in the gauge field ---
in particular the static components of $A^{qu}_0$ which do not commute
with $A^{cl}_0$ ---
alter the relationship
between $\tl$ and $A^{cl}_0$ from that at tree level.
This feeds back into the free energy at two loop order.
As a function of the $\tl$, the free energy is
independent of $\xi$, with
a minimum for $\tl=0$, $F_{two\; loop}(0) \leq F_{two\; loop}(\tl)$.

There are ``unitary'' gauges in which the relationship $A^{cl}_0
= \tl/g$ is unchanged to any loop order.
A reasonable guess is static gauge [\ref{r12}],
imposing $\partial_0 A_0 = 0$.  This is not adequate, however, for static
gauge has a residual gauge freedom of performing static
gauge transformations.  Dividing up
$A_0 = A_0^{cl} + A^{qu}_0$, some
static gauge transformations
rotate the background field $A^{cl}_0$, which is inconsistent.
A gauge which avoids this is static background gauge, obtained by adding
the term
\beq
{\cal L}_{static } \; = \;
\frac{1}{\xi} \; tr \left( \left(
\frac{1}{\xi'}D^{cl}_0 A^{qu}_0 + \partial_i A^{qu}_i \right)^2 \right) \;
\eeq
to the lagrangian.  Letting the gauge parameter $\xi' \rightarrow 0$
fixes $D^{cl}_0 A^{qu}_0 = 0$.
This restricts all fluctuations in
$A^{qu}_0$ to be both static and in directions which commute with
$A^{cl}_0$.  The value of the other gauge fixing parameter $\xi$ is
arbitrary; $\xi$ removes the remaining gauge degeneracy
under static gauge transformations which commute with $A^{cl}_0$.

The calculation of the free energy to one loop order in static background
gauge is illuminating.
At leading order the ghost determinant depends
on the background field as $det(-(D^{cl}_0)^2)$.
In static background gauge the square root of the ghost
determinant equals a product of Vandermonde determinants
for the Wilson line,
\beq
det(D^{cl}_0) \; = \; \prod_x \; |{\cal V}(\tl)|^2
\;\;\; , \;\;\;
{\cal V}(\tl) \; = \; \prod_{i<j} \;
\left(e^{i \tl_i} - e^{i \tl_j}\right) \; .
\eeq
The square of the Vandermonde determinant enters at
each point in space, so
in the action the
logarithm of the product of Vandermonde determinants is proportional
to $\delta^{3}(0) V $, which vanishes in dimensional regularization.
Even in other regularization schemes, however, the
product of Vandermonde determinants from the static background ghosts
do not contribute to the free energy $F(\tl)$.
One factor cancels against the contribution of the spatially longitudinal
mode of the
gluon.\footnote[1]{
This cancellation is similar to that found by Weiss
[\ref{r2}], who imposed the
condition that $A_0$ is static and diagonal
by fiat.  There is no ghost {\it per se}, but
the measure of the functional
integral includes a single factor of the Vandermonde determinant
squared at each point in space.
He obtains the correct free energy at one loop order, where
the product of Vandermonde determinants in the
measure cancels against the contribution of the spatially longitudinal
gluons.
One can check, however, that his method does not give the
correct form of
Gauss' law in the Cartan sub-algebra.
The problem is that the gauge condition
$\partial_0 A_0 = 0$ can only be imposed after
allowing arbitrary variations
in $A_0$, not before.  Imposing
static background gauge as in the text, by
letting $\xi' \rightarrow 0$, avoids this difficulty.
}
The second factor cancels against terms from the
$A_0$ components of the propagator, which give a finite but nonzero
contribution in the limit $\xi' \rightarrow 0$.
After these cancellations the two spatially transverse modes give
the same result as found in covariant gauge.

To turn the path integral of (\ref{e2}) into a partition function we
follow Rossi and Testa [\ref{r8}].
First we transform to
$A_0=0$ gauge.  The gauge transformation which accomplishes this
transformation is the ``partial'' Wilson line, $\Omega(x,\tau)
= {\cal P}exp(i g \int^\tau_0 A_0(x,\tau') d\tau')$.  Many
factors can be ignored in going to
$A_0=0$ gauge.  The Fadeev-Popov determinant
is $det(\partial_0)$, which is independent of temperature and the
background field.  Similarly, when  $0 \leq \tau < \beta$,
all terms in the measure from the gauge
transformation $\Omega(x,\tau)$
can all dropped.
The sole exception is the gauge transformation
for the last slice of euclidean time,
$\Omega(x,\beta)$.
This must be retained because it affects the boundary
conditions: the boundary conditions
are aperiodic,
$A_i(x,\beta) = + A_i^{\Omega(x,\beta)}(x,0)$, while the Wilson
line $L(x) \rightarrow \Omega(x,\beta)$.
Relabeling $\Omega(x,\beta)$ as $\Omega(x)$,
in $A_0=0$ gauge the path integral is
\beq
Z(\tl) \; = \;
\int {\cal D}\Omega(x) \;
\int_{A_i(x,\beta) = + A^{\Omega}_i(x,0)}
\; {\cal D}A_i(x,\tau) \;
\;e^{-S(A_0=0)} \;\;
\delta\left(\tl - \int \frac{d^3 x}{V} \; \lambda(x) \right)
\; .
\label{e3}
\eeq
As before we decompose $\Omega(x) = U(x) \Lambda(x)
U^\dagger(x)$, $\Lambda(x) = e^{i \lambda(x)}$.
In terms of these variables the measure
${\cal D}\Omega = {\cal D}U \; {\cal D}\lambda \; |{\cal V}(\lambda)|^2$,
where the measure naturally includes
the square of the Vandermonde determinant [\ref{r14}].
Thus in $A_0=0$ gauge the constraint on the Wilson line becomes
a constraint on the final gauge transformation, $\Omega(x)$.

An advantage of
$A_0=0$ gauge is that the canonical structure
is manifest [\ref{r8},\ref{r9}].  Thus
the functional integral in (\ref{e3}), written in imaginary
time, is equal to the partition function,
\beq
Z(\tl) \; = \;
\int {\cal D}\Omega(x) \; \sum_{A_i(x)} \;
\langle A_i(x) | \; e^{- \beta {\cal H}} \; | A_i^{\Omega}(x) \rangle
\; \delta\left(\tl - \int \frac{d^3 x}{V} \; \lambda(x) \right)
\; .
\label{e4}
\eeq
All fields $A_i(x,t)$ are functions of a fixed time $t$; this time
is arbitrary, and so the dependence upon it is suppressed.
We assume canonical commutation relations between $A_i(x)$ and
the electric field $E_j(x)$,
$[A^a_i(x),E^b_j(y)] = i \, \delta^{a b} \,
\delta_{i j} \, \delta^3(x-y)$.
${\cal H}$ is the total Hamiltonian, with
the sum over all eigenstates
$|A_i(x)\rangle$; $| A_i^{\Omega}(x) \rangle$
is the gauge transformation of a state under $\Omega(x)$.

To obtain a more useful form of the partition function we introduce
the generators of gauge transformations.
Given $\Omega(x)$, let $\Omega = e^{i \omega}$, and define
the operator
\beq
G(\omega) \; = \; - \; \frac{2}{g} \;
\int d^3 x \; tr\left(\left( D_i \omega(x)\right) E_i(x) \right) \;.
\eeq
The canonical commutation relations imply that
these generators form a representation of the Lie algebra,
\beq
[G\left(\omega_1\right) , G\left(\omega_2\right)]
\; = \; G\left([\omega_1 , \omega_2]\right) \; .
\eeq
By the operator identity $A^{\Omega}_i(x) = e^{- i G(\omega)}
A_i(x)e^{i G(\omega)}$, the gauge transformed state is given by
\beq
|A^{\Omega}_i(x)\rangle \; = \; e^{i G(\omega)} |A_i(x)\rangle \; .
\eeq
Since the operators $G(\omega)$ generate gauge transformations,
they commute with the Hamiltonian, $[G(\omega),{\cal H}] = 0$.

We write the eigenvalues  $\lambda(x) = \tl + \lambda_q(x)$, so
that the constraint becomes $\int d^3 x \, \lambda_q(x) = 0$.
This constraint implies that there is no constant mode in $\lambda_q(x)$,
and is automatically satisfied by requiring that
$\lambda_q(x)$ vanish at spatial infinity.
In this way we trade the constraint for a boundary condition,
\beq
Z(\tl) \; = \;
\int_{\lambda_q(\infty) = 0}
{\cal D}U(x) \; {\cal D}\lambda_q(x)\;
|{\cal V}(\tl + \lambda_q(x))|^2 \; \sum_{A_i(x)} \;
\langle A^{U}_i(x) |
\; e^{- \beta {\cal H} + i G(\tl) + i G(\lambda_q)}
\; | A^{U}_i(x) \rangle
\; ,
\label{e5}
\eeq
To obtain this we start with $|A^\Omega(x)\rangle$,
and by using $\Omega = U \Lambda U^\dagger$, undo the last two
gauge transformations.  $U^\dagger$ is carried through
to change $\langle A_i(x) |$ into $\langle A^U_i(x)|$.  The
generators for the gauge transformations in the Cartan sub-algebra
are left as is.  The factor of $G(\lambda_q(x))$ generates a local,
that for $G(\tl)$ a global, gauge transformation
in the Cartan sub-algebra.
The generator of the global gauge transformation can be written
as a chemical potential for global color charge, ${\cal Q}$:
\beq
G(\tl)
\; = \; 2 \; tr(\tl \, {\cal Q}) \;\;\; , \;\;\;
{\cal Q} \; = \; - \, i \, g \int d^3 x \; [A_i(x),E_i(x)] \; .
\eeq
The appearance of the background field as an imaginary chemical
potential is not surprising: this is obvious at tree level,
using the canonical formalism to expand about a constant field
$A^{cl}_0 = \tl/g$; see, {\it e.g.},
ref.'s [\ref{r3}] and [\ref{r4}].
What is not obvious, and which (\ref{e5}) and
(\ref{e6}) demonstrate,
is how to write the partition function $Z(\tl)$ in a gauge
invariant manner beyond tree level.

It is worth checking that to leading order the partition function
in (\ref{e5}) gives the same free energy as found in the imaginary
time formalism.
The integrations over the gauge transformations
$U(x)$ and $\lambda_q(x)$ can be ignored to this order.
Thus the only change from calculating
the free energy in zero field is the presence
of an imaginary chemical potential for global color charge.
Partition functions of this form have been studied [\ref{r10}];
from these calculations we see that the spatially transverse
gluons give the same free energy as found to one loop order in the
imaginary time formalism.  The spatially longitudinal mode, with
zero energy, cancels the product of Vandermonde determinants
from the measure,
$\prod_x |{\cal V}(\tl + \lambda_q(x))|^2
\approx \prod_x |{\cal V}(\tl)|^2$.

Since $\lambda_q$ vanishes at spatial infinity
we can integrate $G(\lambda_q)$
by parts to obtain
$
G(\lambda_q) = 2 \int d^3 x \; tr(\lambda_q \, D_i E_i)/g \; .
$
This presumes
that the gauge field itself satisfies the appropriate boundary
conditions: $A_i(x) \sim 1/r$ and
$E_i(x) \sim 1/r^2$ as $x \rightarrow \infty$ suffice.  The
only other place where $\lambda_q$ enters is through the
infinite product of
Vandermonde determinants, $\prod_x |{\cal V}(\tl + \lambda(x))|^2$.
As remarked previously, however,
this infinite product vanishes if we adopt
dimensional regularization, which we do.
Then integration over $\lambda_q$ is trivial,
generating a projection operator,
$
{\bf P_C} = \prod_x \delta( tr (t^a_C D_i E_i(x) ) )\, ,
$
where the $t^a_C$ are the $N-1$ elements of the Cartan sub-algebra.
This imposes Gauss' law in the
Cartan sub-algebra at each point in space.  Whence
\beq
Z(\tl) \; = \;
\int {\cal D}U(x) \; \sum_{A_i(x)} \;
\langle A^{U}_i(x) |
\; e^{- \beta {\cal H} \, + \, 2 \, i \, tr(\tl {\cal Q})}
\; {\bf P_C} \; | A^{U}_i(x) \rangle
\; .
\label{e6}
\eeq
This form of the partition function involves only states with positive
norm, the $|A^{U(x)}_i(x)\rangle$, and a positive
measure of integration over $U(x)$.  The background
field enters only through the global color charge,
$e^{2 \, i \, tr(\tl {\cal Q})}$.
While this factor is imaginary, because the sum
always includes states with equal and opposite charge, the total is
necessarily real, so
$e^{2 \, i \, tr(\tl {\cal Q})} \sim cos(2 \, tr(\tl {\cal Q}))$.
Since the cosine function is less than or equal to one,
a nonzero field $\tl$ decreases the partition function,
$Z(\tl) \leq Z(0)$, and so increases the free energy,
\beq
F(0) \; \leq \; F(\tl) \; ,
\label{e7}
\eeq
which concludes our proof.

Our argument is inspired by that of Vafa and Witten [\ref{r15}] for
the $\theta$-angle in $QCD$.
Nonzero $\theta$ appears in the euclidean path integral as a phase
factor, $e^{i \theta {\cal Q}_{top}}$,
where ${\cal Q}_{top}$ is the topological charge.
If the functional measure is positive definite, this implies
that $\theta =0$ and $\pi$ minimize the corresponding free energy.
In our case we need positivity for the sum over states in the partition
function (which is why we take $A_0=0$ gauge) as the background
field appears as a phase factor for global color charge.

A delicate point is the use of dimensional regularization to
eliminate the
infinite product of Vandermonde determinants.  Based upon
calculations to one loop order, where the product of
Vandermonde determinants
cancels in $F(\tl)$, we suggest that this cancellation
persists to arbitrary loop
order.  There is a general reason why it should.  Consider
$Z(\tl)$ in (\ref{e1}),
and then integrate over all $\tl$; the delta function
drops out to give the free energy in zero field.  (Integrating (\ref{e6})
with respect to $\tl$ gives a constraint of zero color charge;
since in infinite volume
the minimum is always at $\tl = 0$, this constraint is
inconsequential.)
But the appropriate
measure for the integration over $\tl$, ${\cal D}\tl \, |{\cal V}(\tl)|^2$,
already includes
a single factor of the
square of the Vandermonde determinant.  Thus
there should not be any additional factors in $Z(\tl)$, infinite or not.

It is trivial to extend the above to include the effects of matter
fields.  Fermion fields, for instance, are added to the
euclidean functional integral,
(\ref{e2}), through antiperiodic boundary conditions.
$\psi(x,\beta) = -\psi(x,0)$.  The above derivations go through
with minor modifications: the states include both gauge fields and
fermions, $|A_i(x),\psi(x)\rangle$, and
the global color charge includes the contribution of the
matter fields as well.  But again we conclude
that $F(0) \leq F(\tl)$.

The partition function also provides insight into
the global \zn symmetry.
If all matter fields lie in the adjoint representation, the
free energy, and so the vacua, are \zn symmetric,
$F(0) = F(\theta_\ell)$ for all $\ell = 0, 1,...(N-1)$.
This \zn symmetry is manifest from the
form of the partition function in (\ref{e6}).
The factor of $e^{2 \, i \, tr(\theta_\ell {\cal Q})} =
e^{i G(\theta_\ell)}$ generates a global \zn gauge transformation on
the states by $\Omega = e^{i \theta_\ell}$.  As
an element of the center of the group this transformation
leaves gluon (and other adjoint) states unaltered,
$e^{2 \, i \, tr(\theta_\ell {\cal Q})} |A^U_i\rangle = |A^U_i\rangle$.

If there are matter fields in the fundamental representation,
the \zn symmetry
is lifted at one loop order: for $\ell \neq 0$,
$F(0) < F(\theta_\ell)$, as the trivial vacuum, $\tl=0$,
is the unique vacuum (modulo the usual factors of $2 \pi$ because
$\tl$ is a phase).
In terms of the partition function, the \zn degeneracy is lifted
because
while states in the adjoint representation
don't change under \zn global gauge
transformations, states in the fundamental representation do:
\beq
e^{2 \, i \, tr(\theta_\ell {\cal Q})} |A^U_i,\psi^U\rangle \; = \;
|A^U_i,e^{i\theta_\ell} \psi^U\rangle \; \neq \;
|A^U_i,\psi^U\rangle \; .
\eeq
In this way, we see that while the Wilson line is defined originally
in euclidean spacetime, its transformation under (aperiodic)
global \zn gauge rotations directly reflects the
transformations of {\it states}, in the partition function,
under global \zn gauge rotations.

We conclude by discussing the nature of states which contribute to
the partition function in (\ref{e6}).  Since by definition $Z(\tl)$
is gauge invariant, the sum in states in (\ref{e6}) must be as well.
A (static)
gauge transformation $V(x)$ changes $A_i \rightarrow A^V_i$; both
the bra and ket states change,
$\langle A^U_i| \rightarrow \langle A^{VU}_i|$ and
$|A^U_i\rangle \rightarrow |A^{VU}_i\rangle \, .$
As the Haar measure is invariant under left multiplication, though,
by $U(x) \rightarrow V^\dagger(x)U(x)$ we can eliminate any such $V(x)$.

While the sum over states is gauge invariant, the states which contribute
are not.  This is very different from the computation of the free energy
in zero field, $F(0)$.  Begin with (\ref{r4}), and just drop the constraint
altogether.  Then $F(0)$ involves states of the form
\beq
| A_i(x) \rangle_{inv} = \int {\cal D}\Omega(x)
\; |A^\Omega_i(x)\rangle \, .
\label{e8}
\eeq
By the properties of the Haar measure,
these states are manifestly gauge invariant [\ref{r1},\ref{r8}],
$|A^V_i\rangle_{inv} = |A_i\rangle_{inv}$.  In
nonzero field, however, the presence of the constraint on the eigenvalues
implies that more than just gauge invariant states contribute to the sum.

Alternately, consider how Gauss' law works.
In zero field it is easy to start from (\ref{e4}) and show that the
integral over $\Omega$ imposes Gauss' law in all directions of the group
[\ref{r1},\ref{r8},\ref{r9}].
This can be seen, albeit less directly,
from (\ref{e6}): for $\tl = 0$, take each state, say the bra
$\langle A^U_i|$, and undo the gauge transformation $U(x)$
by writing it as $\langle A_i| e^{- i G(u)}$,
$U = e^{i u}$.  Doing this for both bra and kets, the
projector for the Cartan sub-algebra becomes
$e^{- i G(u)} {\bf P_C} e^{i G(u)}$.  This is the gauge transformation
of Gauss' law in the Cartan sub-algebra, so integration over the $U(x)$
imposes Gauss' law in all directions of the group algebra.

These manipulations fail in nonzero field, $\tl \neq 0$.
The problem is that the generator
for the local gauge transformation, $G(u)$, need not
commute with that
for the global gauge transformation, $G(\tl)$: there are always
elements $u(x)$ for which $[u(x),\tl] \neq 0$, so $[G(u),G(\tl)]\neq 0$.
Thus in nonzero field, {\it only}
the components of Gauss' law in the Cartan sub-algebra are set to zero,
and not the remaining components.

This feature is generic to {\it any} effective
potential which is a function
of the Wilson line.  Consider, for example, the potential for the real
part of the trace of the Wilson line.  In euclidean spacetime,
\beq
Z_{tr}(J) \; = \;
\int_{A_\mu(x,\beta) = + A_\mu(x,0)} \; {\cal D}A_\mu(x,\tau) \;\;
e^{-S - J \int d^3 x \, Re(tr \, L )} \;\;
\; ,
\label{e10}
\eeq
using a source $J$ coupled to the trace of the Wilson line.
Using the above techniques this equals the partition function
\beq
Z_{tr}(J) = \int {\cal D}\Omega(x)\, \sum_{A_i} \,
\langle A_i(x)| \;
e^{- \beta {\cal H} - J \int d^3 x \, Re(tr \,\Omega) } \;
| A_i^{\Omega}(x)\rangle \, .
\label{e11}
\eeq
Again, while the sum over states is gauge invariant, the individual
states are not: the source dependent state
\beq
|A_i(x), J\rangle
= \int {\cal D}\Omega(x) \;
e^{ - J \int d^3 x \, Re(tr \, \Omega) } |A^\Omega_i(x)\rangle
\label{e12}
\eeq
is not gauge invariant: $|A^V_i, J\rangle \neq |A_i, J\rangle$ when
$J \neq 0$.

This behavior is not entirely
unexpected.  After all, the vacuum expectation value of the trace
of the Wilson line,
$\langle tr \, L(x) \rangle = \partial ln(Z(J))/\partial J(x)|_{J=0}$, is
the wave function for an infinitely heavy test quark [\ref{r1}].  In
terms of Gauss' law, a single Wilson line corresponds
to the insertion of a point source of unit
strength at that point [\ref{r1}].
A source for the (trace of the) Wilson line, spread out over all
of space, then modifies Gauss' law everywhere.  From (\ref{e11}),
when $J(x) \neq 0$ no component of Gauss' law vanishes, not even
those in the Cartan sub-algebra.  Amusingly, the right hand
side of Gauss' law, which is proportional to the source $J$, is only
real if $J$ is purely imaginary.

Partition functions similar to that in
(\ref{e6}) have been studied
before [\ref{r10}].  The difference is that these works all assume
that only gauge invariant states, as in (\ref{e8}), contribute:
\beq
Z_{inv}(\tl) \; = \;
\sum_{A_i(x)} \;
\langle A_i(x) |_{inv}
\; e^{- \beta {\cal H} + 2 \, i \, tr(\tl {\cal Q})}
| A_i(x) \rangle_{inv}
\; ,
\label{e13}
\eeq
At zero field everything reduces to the standard
free energy: $Z(0) = Z_{inv}(0)$.
Since $Z_{inv}$ contains
no term in the measure for the Vandermonde determinant,
at one loop order $Z(\tl)$ and $Z_{inv}(\tl)$ differ by a product
of Vandermonde determinants, as the spatially longitudinal gluons give
$1/\prod_x |{\cal V}(\tl)|^2$.  This leads one to suspect that perhaps
to arbitrary loop order, $Z(\tl)$ and $Z_{inv}(\tl)$ differ only
by terms which vanish in dimensional regularization.

Nevertheless, (\ref{e13}) has one serious drawback: in euclidean
spactime it does not
correspond to the functional integral for a gauge invariant
potential.  Using the definition of the
invariant state $|A_i(x)\rangle_{inv}$,
\beq
Z_{inv}(\tl) \; = \; \int {\cal D}\Omega_1(x) \;
\int {\cal D}\Omega_2(x) \;
\sum_{A_i(x)} \;
\langle A_i(x) |
\; e^{- \beta {\cal H}}
| A^{\Omega_f}_i(x) \rangle \;\;\; , \;\;\;
\Omega_f(x) \; = \; \Omega_1(x) e^{i\tl} \Omega^\dagger_2(x) \; .
\label{e20}
\eeq
This is similar to the partition function in (\ref{e4}),
with the replacement of the gauge transformation $\Omega$ by
$\Omega_f$.  In (\ref{e4}) the background
field $\tl$ is related to a gauge invariant quantity, the spatial
average over the phases of the eigenvalues of $\Omega(x)$.
In (\ref{e20}), however, $\tl$ is not related to anything
gauge invariant, since the eigenvalues of
$\Omega_f = \Omega_1 e^{i \tl} \Omega^\dagger_2$ are
not $e^{i \tl}$.  Computing the euclidean path
integral which corresponds to (\ref{e20}) [\ref{r1}] leads to an
expansion about constant $A^{cl}_0 = \tl/g$, which is
gauge dependent beyond leading order.

Consider the consequences if, by
dint of prejudice, one
insists that only gauge invariant states should
contribute to the partition
functions of gauge theories.
Then construct the effective potential
for the trace of the Wilson line, as in (\ref{e11}),
everywhere replacing the states $|A_i\rangle$ with gauge invariant
states, $|A_i\rangle_{inv}$.
But as $|A^\Omega_i\rangle_{inv} = |A_i\rangle_{inv}$,
this partition function gives a
{\it trivial} potential for the trace of
the Wilson line, $Z_{inv}(J) = \prod_x
\int {\cal D}\Omega(x) e^{- J \int d^3 x \,
Re( tr \, \Omega)}$: this
potential is independent of temperature or the field content,
and as an infinite product, vanishes in dimensional regularization.
This is untenable, manifestly in contradiction with numerical
simulations of euclidean lattice gauge theories.

Partition functions as in (\ref{e6})
do produce novel thermodynamic behavior.
In ordinary partition functions
each state makes a positive contribution to $Z$, which implies that
the pressure is positive.
In nonzero field, however,
as the sum over states includes a factor of
$cos(2 \, tr(\tl {\cal Q}))$, states with nonzero charge can make a
negative contribution to $Z(\tl)$, and consequently the pressure
can be negative.
For instance, even for a single scalar field coupled to an abelian
gauge field, there are regions of $\tl$ in which the
pressure is negative.

Dixit and Ogilvie [\ref{r3}]
observed that in theories with dynamical fermions,
while \zn transforms of the usual vacuum have greater
free energy than the trivial vacuum, they are often
metastable.  The cosmological implications of such \zn ``bubbles''
were developed by Ignatius, Kajantie, and Rummukainen [\ref{r2}].
In ref. [\ref{r4}], however, it was pointed out that the pressure of the
metastable \zn state can be negative, depending upon matter content
of the theory.  This includes the physically interesting case of three
colors and six massless
flavors, as applies at temperatures above the restoration of the
$SU(2) \times U(1)$ symmetry.
While the pressure of $W$, $Z$, and Higgs bosons acts to
make the total pressure positive, it is rather disquieting to find that
the pressure of a subsystem is negative.
On these, and other grounds,
the authors of [\ref{r4}] argue that the metastable
\zn states are not thermodynamically accessible.

We suggest the contrary.  Surely a system with negative pressure
is not thermodynamically stable; indeed, it will not even be metastable
for long.
But we propose that if a path integral is gauge invariant and well defined
in euclidean spacetime, then the corresponding partition function,
as in (\ref{e6}) and (\ref{e11}), {\it do}
represent physically realizable systems.  It's just that
there's a richer variety of partition functions
possible in gauge theories.

We thank M. Creutz, S. Gavin, C. Korthals-Altes, B. Svetitsky,
and L. Trueman for
helpful discussions.  This research
was supported in part by the U.S. Department of Energy under
contract DE--AC02--76CH00016.

\noindent{\bf References}
\newcounter{nom}
\begin{list}{[\arabic{nom}]}{\usecounter{nom}}
\item
A. M. Polyakov, \phl{72B}{477}{78};
L. Susskind, \prd{20}{2610}{79};
B. Svetitsky, Phys. Rep. {\bf 132}, 1 (1986).
\label{r1}
\item
N. Batakis and G. Lazarides, \prd{18}{4710}{78};
D. Gross, R.D. Pisarski, and L.G. Yaffe, \rmp{53}{43}{81},
especially sec.'s III.B, V.A, and appendix D;
N. Weiss, \prd{24}{475}{81}; \ibid{25}{2667}{82}.
\label{r2}
%
%
\item
J. Polonyi and K. Szlachanyi, \phl{110B}{395}{82};
Y. Hosotani, \anp{190}{233}{89};
J. Bohacik, \prd{42}{3554}{90};
C.-L. Ho and Y. Hosotani, \nup{B345}{445}{90}
V. Dixit and M.C. Ogilvie, \phl{B269}{353}{91};
J. Ignatius, K. Kajantie, and K. Rummukainen, \prl{68}{737}{92};
J. Polchinski, \ibid{68}{1267}{92};
S. Chapman, LBL preprint LBL-33116 (July, 1992).
\label{r3}
\item
V. M. Belyaev, I. I. Kogan, G. W. Semenoff, and N. Weiss,
\phl{B277}{331}{92};
W. Chen, M. I. Dobroliubov, and G. W. Semenoff, \prd{46}{R1223}{92}.
\label{r4}
%
%
\item
R. Anishetty, \jph{G10}{423}{84}; \ibid{G10}{439}{84};
K. J. Dahlem, \zpc{C29}{553}{85};
J. Polonyi, \nup{A461}{279}{87};
S. Nadkarni, \prd{34}{3904}{86}; \ibid{38}{3287}{88};\prl{60}{491}{88};
V. M. Belyaev, \phl{B241}{91}{90};
V. M. Belyaev and V. L. Eletsky, \zpc{45}{355}{90};
K. Enqvist and K. Kajantie, \ibid{47}{291}{90};
J. Polonyi and S. Vazquez; \phl{B240}{183}{90};
M. Burgess and D. J. Toms, \anp{210}{438}{91};
M. Oleszczuk and J. Polonyi, MIT preprints MIT-CTP-1984 (June, 1991),
Regensburg preprints  TPR-92-33 and TPR-92-34;
K. Shiraishi and S. Hirenzaki, \zpc{53}{91}{92}.
\label{r5}
\item
V. M. Belyaev, \phl{B254}{153}{91}.
\label{r6}
\item
T. Bhattacharya, A. Gocksch, C. P. Korthals Altes and R. D. Pisarski,
\prl{66}{998}{91}; \nup{B383}{497}{92}.
\label{r7}
\item
G. C. Rossi and M. Testa, \nup{B163}{109}{80}; \ibid{B176}{477}{80};
\ibid{B237}{442}{84}.
\label{r8}
\item
Gauge fields: Introduction to
Quantum Theory, L. D. Fadeev and A. A. Slavnov (Benjamin/Cummings,
Reading, 1980).
\label{r9}
\item
K. Redlich and L. Turko, \zpc{5}{201}{80};
L. Turko, \phl{104B}{153}{81};
C. DeTar and L. McLerran, \ibid{119B}{171}{82};
M. I. Gorenstein, S. I. Lipskikh, V. K. Petrov, and G. M. Zinovjev,
\ibid{123B}{437}{83};
M. I. Gorenstein, O. A. Mogilevsky, V. K. Petrov, and G. M. Zinovjev,
\zpc{18}{13}{83};
H.-Th. Elze, W. Greiner, and J. Rafelski, \phl{124B}{515}{83};
B.-S. Skagerstam, \ibid{133B}{419}{83}; \zpc{24}{97}{84};
P. A. Amundsen and B.-S. Skagerstam, \phl{165B}{375}{85};
L. D. McLerran and A. Sen, \prd{32}{2794}{85};
H.-Th. Elze and W. Greiner, Phys. Rev. {\bf A 33}, 1879 (1986);
S. I. Azakov, P. Salomonson, and B.-S. Skagerstam, \prd{36}{2137}{87};
D. E. Miller and K. Redlich, \prd{35}{2524}{87}; \ibid{37}{3716}{88};
H.-Th. Elze, D. E. Miller, and K. Redlich, \ibid{35}{748}{87};
A. Le Yaouanc, L. Oliver, O. Pene, J.-C. Raynal, M. Jarfi, and
O. Lazrak, \ibid{37}{3691}{88}; \ibid{37}{3702}{88}; \ibid{38}{3256}{88};
\ibid{39}{924}{89};
S. P. Klevansky and R. H. Lemmer, \ibid{38}{3559}{88}.
\label{r10}
\item
L. O'Raifeartaigh, A. Wipf, and H. Yoneyama, \nup{B271}{653}{86},
and references therein.
\label{r11}
\item
A. Burnel, R. Kobes, G. Kunstatter, and K. Mak, \anp{204}{247}{90},
and references therein.
\label{r12}
\item
The Classical Groups, H. Weyl
(Princeton University Press, Princeton, 1946).
\label{r14}
\item
C. Vafa and E. Witten, \prl{53}{535}{84}.
\label{r15}
\end{list}
\end{document}